# Estimating the economic value of ultrafine particles information:

# A contingent valuation method

Eunjung Cho[a] and Youngsang Cho[a*]

[a] Department of Industrial Engineering, College of Engineering, Yonsei University, 50 Yonsei-Ro, Seodaemun-gu, Seoul, 03722, South Korea

Email: EC, ejung720@yonsei.ac.kr; YC, y.cho@yonsei.ac.kr

[*]**Corresponding author:**

Youngsang Cho, PhD

Professor

Tel: +82-2-2123-5727

Fax: +82-2-364-7807

E-mail: y.cho@yonsei.ac.kr

# Estimating the economic value of ultrafine particles information:

# A contingent valuation method

**Abstract**

Global concern regarding ultrafine particles (UFPs), which are particulate matter (PM) with a diameter of less than 100nm, is increasing. These particles—with more serious health effects than PM less than 2.5µm (PM2.5)—are difficult to measure using the current methods because their characteristics are different from those of other air pollutants. Therefore, a new monitoring system is required to obtain accurate UFPs information, which will raise the financial burden of the government and people. In this study, we estimated the economic value of UFPs information by evaluating the willingness-to-pay (WTP) for the UFPs monitoring and reporting system. We used the contingent valuation method (CVM) and the one-and-one-half-bounded dichotomous choice (OOHBDC) spike model. We analyzed how the respondents' socio-economic variables, as well as their cognition level of PM, affected their WTP. Therefore, we collected WTP data of 1,040 Korean respondents through an online survey. The estimated mean WTP for building a UFPs monitoring and reporting system is KRW 6,958.55-7,222.55 (USD 6.22–6.45) per household per year. We found that people satisfied with the current air pollutant information, and generally possessing relatively greater knowledge of UFPs, have higher WTP for a UFPs monitoring and reporting system. The results can be used to establish new policies response to PM including UFPs.

# 1. Introduction

Numerous epidemiological studies have been conducted on the effects of particulate matter (PM) with a diameter of less than 2.5µm (PM2.5) on human health (Dockery et al., 1993; Hart et al., 2015; Li et al., 2018). The literature, consequently, is replete with evidence of its negative effects on human beings (Atkinson et al., 2014; Burgan et al., 2010; Cesaroni et al., 2014; Yuan et al., 2019).

However, during recent years, general public concern—and interest—regarding ultrafine particles (UFPs), which are PM with a diameter of less than 100nm, has also increased. The characteristics of PM depend on its size (Morawska et al., 2008), and there are four distinguishing characteristics of UFPs. First, they constitute less than 20% of the total mass concentration of particles, but more than 90% of the total number concentration of particles, compared to PM with a diameter of less than 10µm (PM10) and PM2.5 (Kittelson, 1998; Kumar et al., 2009). Second, UFPs have a high share in direct emissions from anthropogenic sources, such as road transportation and power plants, whereas PM2.5 has a high share in secondary sources, that is, through chemical processes in the atmosphere (Kittelson, 1998; Liang et al., 2016; Morawska et al., 2008). UFPs emitted from road transportation account for over 60% of the total air pollution, as compared to non-road transportation (19%) and domestic combustion (13%) (Kumar et al., 2014). Third, various natural factors, such as wind direction, wind speed, and breathability, affect UFPs concentration (Buccolieri et al., 2010; Chen et al., 2016). In urban areas, the vortex effect caused by dense traffic concentration and the temperature gradient (Kumar et al., 2008; Marini et al., 2015), results in UFPs—emitted by vehicular fuel combustion—hanging in the atmosphere for a long time, at high concentration levels. Kumar et al. (2014) compared UFPs concentration in Asian countries, such as China and India, and European countries. They found that outdoor average UFPs concentration in Asian cities is about four-times higher than in European cities. According to them, understanding the variability of UFPs is the key to designing effective monitoring strategies and estimating the relationships between UFPs and health effects in urban areas. Fourth, UFPs have low mobility. Choi et al. (2018) built a mobile monitoring platform to measure UFPs at intervals of 90m before and after the intersection center, and found that their concentration peaks within 30m of the intersection, decreasing sharply thereafter.

Similar to PM10 and PM2.5 studies, there are studies on health effects of UFPs exposure. These particles are too small for the human nose and bronchioles to effectively filter out, resulting in their deep absorption into the alveoli or the membranes (HEI, 2013). Thus emanate numerous respiratory diseases, and the smaller the particle size, the greater the risk (Penttinen et al., 2001). According to Chen et al. (2016), about 50% of all PM deposited on the alveoli is of the size of 20nm, and 10–20% from 100nm to 2.5µm. Stafoggia et al. (2017) analyzed the relationship between short-term exposure to UFPs concentration and mortality in eight European countries. The results showed a 0.35% increase in non-accidental mortality as the number of UFPs increased by 10,000 particles/m$^3$. Liu et al. (2018) estimated the relationship between maximum blood pressure, minimum blood pressure, high sensitivity-C-reactive protein, and UFPs concentrations for 100 non-smokers in Taiwan. Consequently, when the UFPs concentration increased by 0.97µg/m$^3$, maximum blood pressure, minimum blood pressure, and high sensitivity-C-reactive protein increased by 6.3%, 5.6%, and 8.5%, respectively—higher than the effects of PM10 and PM2.5. In Korea, Song et al. (2011) analyzed the effect of UFPs concentration on respiratory system functions in 41 children aged 8–12 years with atopic dermatitis. They found a 3.1% increase in itch symptom score in children with atopic dermatitis as the number concentrations of UFPs increased by the interquartile range. Although a few studies suggest the high potential of adverse health effects due to UFPs exposure, more studies are required to prove the same with greater reliability (Ohlwein et al., 2019; Schraufnagel, 2020).

The current PM response policies and measurement methods cannot properly reflect the characteristics of UFPs; therefore, they need to be managed separately from existing air pollutants. Consequently, several developed countries have included UFPs in their inventory of fine PM (Lewis et al., 2018). However, several other countries, including Korea, do not have in place adequate standards and regulatory limits for UFPs. Therefore, establishment of a future policy on air pollutants, including UFPs, necessitates a wide range of UFPs studies to collect scientific evidence. Additionally, prevention of potential UFPs risks requires disclosure of information to the public so as to facilitate voluntary avoidance of UFPs exposure. To collect and disseminate this information, a new UFPs monitoring and reporting system—different from traditional systems—is required, which will raise the financial burden on the government and people. Moreover, research on public perception of UFPs and the economic value of its information remains insufficient.

This study has two main aims: to analyze the economic value of UFPs information by using the contingent valuation method (CVM) to estimate the willingness-to-pay (WTP) for building a UFPs monitoring and reporting system; and, to derive policy implications for the government to respond to UFPs. The remainder of this paper is organized as follows: Section 2 explains the model and data used in this study. Section 3 presents estimation results and discusses implications. Section 4 provides the conclusions and limitations of this study.

## 2. Methods and Materials

### 2.1 Methods

This study used the CVM to estimate the WTP for building a UFPs monitoring and reporting system, to evaluate the economic value of UFPs information. The CVM has been widely used for economic valuation of non-market goods, especially environmental goods, and consumer perception analysis (Bateman and Langford, 1997; Han et al., 2011; Mwebaze et al., 2018). To estimate the WTP for non-market goods using the CVM, it is important to design a questionnaire that facilitates respondents' comprehension of issues and obtains their WTP information. Generally, payment card, open-ended, and dichotomous choice (DC) formats are employed to elicit respondents' WTP data through the CVM. In the DC format, pre-set bids are presented to the respondents with a "yes" or "no" choice, and econometric analysis—the probit or logit model—is used to estimate the WTP. We designed the questionnaire using the DC format because of its following strengths: easy to answer, low biases, and low likelihood of estimating unreliable WTP (Hanemann, 1984; Oerlemans et al., 2016).

The DC format is divided into the single-bounded dichotomous choice (SBDC) that asks a single bid (Bishop and Heberlein, 1979), and the double-bounded dichotomous choice (DBDC) that asks once more for double or half of the first bid, depending on the respondent's first answer (Hanemann, 1985). The SBDC has a relatively low non-response rate and is convenient because it asks each respondent only one bid; however, it has the disadvantage of being statistically inefficient. The DBDC, in contrast, has high efficiency, but one disadvantage: the first and second bids are not independent, which may cause bias. To overcome SBDC's inefficiency and DBDC's bias, Cooper et al. (2002) suggested the one-and-one-half-bounded dichotomous choice (OOHBDC), in which either the lower or upper bid is randomly given to each respondent as an initial bid. If respondents answer "no" to the lower bid or "yes" to the upper bid, the survey ends. However, if they answer "yes" to the lower bid, then the upper bid is presented as the second question, and if they answer "no" to the upper bid, then the lower bid is presented as the second question. This study used the OOHBDC method to reduce SBDC's inefficiency and DBDC's bias. In the OOHBDC method, if respondents answer "no" to the lower bid or "no-no" to the upper bid, then their WTP may be zero or between zero and the lower bid. It is important to identify whether the respondent's WTP is actually zero or between zero and the lower bid, so as to estimate their accurate WTP; we used the spike model to solve this problem (Kriström, 1997). The OOHBDC spike model asks an additional question if respondents answer "no" to the lower bid or "no-no" to the upper bid: "Are you willing to pay at least KRW 1 to build the UFPs monitoring and reporting system?"

This study applied the utility difference model suggested by Hanemann (1984) to estimate the WTP using the OOHBDC spike model. Each respondent has the indirect utility function $u(j,m;S)$, where $j$ is the status of the UFPs monitoring and reporting system: when $j$ is equal to 1, it means the UFPs monitoring and reporting system is presented, otherwise $j$ is 0. $m$ is the respondent's income, and $S$ is the vector of respondent's socio-economic and cognition characteristics. The indirect utility function can be expressed as the observable deterministic part, $v(j,m;S)$, and unobservable stochastic part, $\varepsilon_j$, as follows:

$$u(j,m;S) = v(j,m;S) + \varepsilon_j, \tag{1}$$

where $\varepsilon_j$ is an independently and identically distributed (i.i.d.) variable with a zero mean.

When the respondent answers "yes" to the question "Are you willing to pay $A$ to build the UFPs monitoring and reporting system?," to maximize his/her utility, the probability of answering "yes" is expressed as follows:

$$\Pr\{"yes"\} = \Pr\{\Delta v(A) \geq \eta\} \equiv F_\eta[\Delta v(A)], \tag{2}$$

where $\eta$ is the difference of error terms, $\varepsilon_0 - \varepsilon_1$, and $F_\eta(\cdot)$ is the cumulative distribution function (CDF) of $\eta$. Meanwhile, if the WTP (denoted as $W$) of the respondent is greater than or equal to $A$, the respondent will answer "yes," otherwise "no." The probability that respondent answers "yes" can also be expressed as follows:

$$\Pr\{"yes"\} = \Pr\{W \geq A\} \equiv 1 - G_W(A), \tag{3}$$

where $G_W(A)$ is CDF of $W$. When we consider equations (2) and (3) together, we can derive $1 - G_W(A) \equiv F_\eta[\Delta v(A)]$.

We assume that $A_i$ is an initial bid presented to the respondent $i$, and $A_i^L$ and $A_i^U$ represent lower and upper initial bids, respectively. There are eight possible outcomes in the OOHBDC spike model:

$$I(A_i^U) = \begin{cases} I_i^Y = 1 & (\text{respondent } i \text{ answers "yes"}), \text{where } A_i^U < W < \infty \\ I_i^{NY} = 1 & (\text{respondent } i \text{ answers "no-yes"}), \text{where } A_i^L < W < A_i^U \\ I_i^{NNY} = 1 & (\text{respondent } i \text{ answers "no-no-yes"}), \text{where } 0 < W < A_i^L \\ I_i^{NNN} = 1 & (\text{respondent } i \text{ answers "no-no-no"}), \text{where } W = 0 \end{cases}, \tag{4}$$

and

$$I(A_i^L) = \begin{cases} I_i^{YY} = 1 & (\text{respondent } i \text{ answers "yes-yes"}), \text{where } A_i^U < W < \infty \\ I_i^{YN} = 1 & (\text{respondent } i \text{ answers "yes-no"}), \text{where } A_i^L < W < A_i^U \\ I_i^{NY} = 1 & (\text{respondent } i \text{ answers "no-yes"}), \text{where } 0 < W < A_i^L \\ I_i^{NN} = 1 & (\text{respondent } i \text{ answers "no-no"}), \text{where } W = 0 \end{cases}, \tag{5}$$

where the indicator function $I(\cdot)$ has a value of 1 if the proposition is true; otherwise, it is 0. Using eight indicator functions, the log-likelihood function for the OOHBDC spike model is expressed as follows:

$$\ln L = \sum_{i=1}^{N} \left\{ \begin{array}{l} (I_i^Y + I_i^{YY})\ln[1 - G_W(A_i^U)] + (I_i^{YN} + I_i^{NY})\ln[G_W(A_i^U) - G_W(A_i^L)] \\ + (I_i^{NNY} + I_i^{NY})\ln[G_W(A_i^L) - G_W(0)] + (I_i^{NNN} + I_i^{NN})\ln[G_W(0)] \end{array} \right\}. \tag{6}$$

Assuming that the respondent's WTP has a logistic CDF, the spike model of $G_W(A)$ with parameter $a$, $b$ is:

$$G_W(A) = \begin{cases} [1 + \exp(a - bA)]^{-1} & \text{if } A > 0 \\ [1 + \exp(a)]^{-1} & \text{if } A = 0 \\ 0 & \text{if } A < 0 \end{cases}. \tag{7}$$

Here, the spike is defined as $[1+\exp(a)]^{-1}$ and the mean WTP is calculated as $(1/b)\ln[1+\exp(a)]$.

## 2.2  Survey design and data

We designed the survey to estimate the WTP for building a UFPs monitoring and reporting system. Initially, we planned a face-to-face survey; however, it was difficult to meet the respondents owing to COVID-19. Therefore, the survey was conducted online. At the beginning of the questionnaire, we explained the definition and characteristics of UFPs, and the differences between UFPs and PM2.5. Thereafter, we described the kinds of information that the UFPs monitoring and reporting system will provide to the public: (1) one-hour average data, presented online in real-time; (2) monthly/annual reports; and (3) predicted UFPs concentration data and warning alert issuance in case of high concentration. We also explained how to use UFPs information, based on policy suggestions of previous studies (Choi et al., 2018; Choi et al., 2012; Hu et al., 2009; Lewis et al., 2018).

In this survey, we selected an additional payment of income tax as the payment vehicle for building the UFPs monitoring and reporting system. Additionally, we explained to the respondents that they would have to pay over the next five years, considering 90% of PM2.5 monitoring stations in Korea are installed for five-year terms.[1] Consequently, we presented the question: "Are you willing to pay an additional KRW [bid] each year for the next five years from the income tax paid by your household to build the UFPs monitoring and reporting system?"

We conducted a pilot survey of 455 respondents to determine the initial bid sets to be presented to respondents in the actual survey. Based on the results of the pilot survey, we excluded the upper and lower 5% responses to remove the bias, considering them as outliers. Then, we determined 10 sets of initial bids for the actual survey: KRW (1,000; 2,000), (2,000; 3,000), (3,000; 4,000), (4,000; 5,000), (5,000; 7,000), (7,000; 9,000), (9,000; 11,000), (11,000; 14,000), (14,000; 17,000), (17,000, 20,000).[2] The survey was conducted by Gallop Korea, a specialized market research company, from February 4–9, 2021. Applying stratified sampling, the company selected 1,042 respondents between 20–69 years of age, and completed the online survey. We excluded two respondents who did not answer the additional questions, and used a total of 1,040 respondents in the analysis.

The characteristics of the respondents in the context of the total Korean population are described in Table 1. The proportions of sex, age, region, and average monthly income per household are similar between survey respondents and the total population. However, the education level of the respondents is relatively higher than the general population—this could be due to the inclusion of internet savvy individuals in the survey, who generally have a high level of education (Ünver, 2014). We compared the average monthly expense for anti-dust products with another survey sample of Min (2019) in Table 1. The proportion of respondents

---

[1] Korea began installing PM2.5 monitoring stations from 2015, and there were a total of 458 stations as of October 2020 (ME, 2020a). Of these, about 90% (400 stations) were installed by December 2019 (ME, 2020b).

[2] USD 1.0 = KRW 1,119.1 at the time of conducting this study (2021/02/09) (Source: IMF, 2021)

who spend over KRW 50,000 was lower, but the level of spending under KRW 50,000 was higher. This reflected the increase in the sale of face mask—a representative anti-dust product—due to COVID-19 after 2019 (Wu et al., 2020).

Insert [Table 1] about here

We presented four additional questions to confirm the cognition level of respondents regarding PM: seriousness, adverse health effect, reliability of current real-time PM data, and perception of UFPs. Table 2 shows the results for the 1,040 respondents. Most people are sensitive to PM and 932 people (89.62%) responded that the PM problem in Korea is serious. Additionally, 518 people (49.81%) responded that the real-time PM data currently provided by the government is reliable. However, people's perception of UFPs is low, whereas their interest in PM is considerably high.

Insert [Table 2] about here

During the survey, 20 sets of bids were allocated to 1,040 respondents in a similar ratio. Table 3 presents the distribution of respondents by initial bid and their answers. A total of 118 people (21.7%) responded "no-no-no" to the upper initial bid, and 131 people (26.5%) responded "no-no" to the lower initial bid. From this, we found that 249 out of all respondents (23.9%) have zero WTP for building a UFPs monitoring and reporting system.

Insert [Table 3] about here

Respondents' intention for zero WTP can either be a case of their WTP being true zero, or a protest response to the survey. A protest response is observed when respondents are dissatisfied with the questionnaire or think there is insufficient information about the survey issue. In this case, their WTP may not be true zero, and, therefore, these answers are most probably a protest response to the survey. In this survey, we presented an additional question to zero WTP respondents to ascertain the reason for their zero WTP, so we could distinguish the protest WTP respondents. Table 4 shows the reasons mentioned for zero WTP. We can consider the respondents who responded with "Not enough information to judge" as protest WTP respondents (Tolunay and Başsüllü, 2015). Previous studies suggested two methods for treating protest WTP respondents. First, these respondents were excluded from the CVM analysis because their responses did not reflect their actual preferences (Jorgensen and Syme, 2000; Meyerhoff and Liebe, 2010). Second, they were included in the data set because their exclusion from the analysis, could have overestimated the mean WTP (Fonta et al., 2010; Strazzera et al., 2003). In this study, there were 38 (3.65%) protest WTP respondents, and we included their WTP as zero in the basic analysis to avoid the overestimation bias; we also presented the estimated WTP, excluding the protest WTP respondents for comparison.

Insert [Table 4] about here

## 3. Results and Discussion

Table 5 shows the estimation results of three OOHBDC spike models: Model 1 without any covariate; Model 2 with socio-economic covariates, such as sex, age, household income, and anti-dust expense; and Model 3 with cognition covariates as well as socio-economic covariates. The estimated WTP can be interpreted as the household's WTP because we presented the question based on the "income tax paid by respondent's household" in the survey. The Wald test was used to test if the coefficients' values in the estimated equation were zero. According to the results, this hypothesis is rejected at the 1% significant level in all three models.

Insert [Table 5] about here

First, in the Model 1 results, all estimates are statistically significant at the 1% level. The estimated mean WTP is KRW 7,222.55 (USD 6.45) per household per year and is statistically significant at the 1% level. Second, in Model 2 results, sex and income are not statistically significant. In contrast, age and anti-dust expense are statistically significant at the 1% level and the estimated covariates of these variables are positive. This indicates that respondents who are older or spend more money on anti-dust expenses are willing to pay more for building a UFPs monitoring and reporting system. The estimated mean WTP derived from Model 2 is KRW 7,196.33 (USD 6.43) per household per year and statistically significant at the 1% level. Third, in the results of Model 3, age, seriousness, and adverse health effects are statistically significant at the 5% level and the reliability of real-time PM data and perception of UFPs are statistically significant at the 1% level. The estimated coefficients of these variables are positive, which indicates that respondents who are older, who take PM concentration in Korea seriously, who think that PM has a negative impact on their health, or who possess deeper background knowledge of UFPs, are willing to pay more.

Particularly, the coefficient of the reliability of real-time PM data is the highest, which can be interpreted as following: the more reliable the PM information provided by the government, the greater the WTP for building a UFPs monitoring and reporting system. The estimated mean WTP is KRW 6,958.55 (USD 6.22) per household per year and statistically significant at the 1% level.[3]

Additionally, we conducted the Monte Carlo simulation to calculate the 95% and 99% confidence intervals of the estimated WTP (Krinsky and Robb, 1986; Lee and Cho, 2020). Based on the estimated coefficients and their variance-covariance matrix, we generated 5,000 replications and calculated the mean WTP. Thereafter, we omitted 2.5% and 0.5% at both ends to obtain the 95% and 99% confidence intervals, respectively. The last two rows of Table 5 present the 95% and 99% confidence intervals of WTP, respectively, derived from the Monte Carlo simulation.

Expanding the household-level WTP to the total population could be helpful in estimating the national level economic value of UFPs information. As the total number of households was 23,093,108 in 2020 (MOIS, 2021), the national level annual WTP is KRW 166.79 billion (USD 148.95 million). We presented five years for the payment period in our survey scenario, so the total WTP to receive UFPs information nationwide is approximately KRW 833.96 billion (USD 744.75 million).

To the best of our knowledge, there are no studies on the economic value of UFPs information, but a few studies have estimated the economic value of air quality improvement. Table 6 shows previous studies that used the CVM to estimate the WTP for air quality improvement. The different scenarios used in previous studies to explain specific methods for air quality improvement include strengthening policies (Kim et al., 2018; Wang and Zhang, 2009), reducing air pollutants (Akhtar et al., 2017; Wang et al., 2006), and reducing the risk of mortality due to air pollution (Istamto et al., 2014; Lee et al., 2011; Ligus, 2018; Sun et al., 2016; Vlachokostas et al., 2011).

---

[3] Excluding protest WTP respondents, mean WTP were KRW 7,459.89, 7,429.49, and 7,136.52 for the three models, respectively. These estimated WTP were increased by 3.29%, 3.24%, and 2.56%, respectively, compared to the results including protest WTP respondents.

Direct comparison of previous studies' WTP with our result is difficult because there are differences as regards survey year, scenarios, payment vehicle, and survey unit (per household, per person). However, in general, the WTP derived from the scenario of avoiding death, estimated in Western countries, is higher than those derived from the scenario of improving air quality, estimated in Asian countries. In terms of payment vehicle, studies analyzing the WTP for air quality improvement mainly selected tax increases and voluntary payments. Kim et al. (2018) estimated the WTP for strengthening the PM2.5 concentration reduction policy in Korea using the OOHBDC. They asked respondents about their WTP for policy enhancement, such as tightening the overall regulation of the PM2.5 sources, expanding the concentration monitoring station, and improving the management of deteriorated diesel vehicles. They used an annual income tax increase as the payment vehicle, and found that the mean WTP for the enforcement of the PM2.5 concentration reduction policy was USD 4.97 (USD 5.09 in 2020) per household per year. Wang and Zhang (2009) estimated the WTP for improving air quality through enforcing a strict national air quality standard in Jinan—a Chinese city with the poorest air quality. They asked respondents how much they were willing to pay voluntarily and found that the WTP was CNY 100 (USD 22.29 in 2020) per person per year.

Insert [Table 6] about here

## 4. Conclusions

The negative health effects of PM have been demonstrated by several studies, and Korea—like numerous other countries—allocates enormous annual budgets to reduce PM emission and improve air quality. Korea's Ministry of Environment (ME)'s 2021 budget for atmospheric environment is KRW 2,922.71 trillion (USD 2.60 trillion) (ME, 2021). Of late, concerns about UFPs beyond PM2.5 have increased in Korea; however, studies that can reliably demonstrate their potential risks are insufficient. Moreover, the current measurement methods are not suitable for UFPs that have characteristics different from other air pollutants. Obtaining accurate information about UFPs is necessary to conduct studies on their emission sources, characteristics, and health effects. Therefore, we estimated the economic value of UFPs information using the CVM, and the scenario was building a new UFPs monitoring and reporting system.

The mean WTP was estimated as KRW 6,958.55-7,222.55 (USD 6.22-6.45) per household per year, applying three models. We identified that the cognitive level regarding PM rather than income had a positive effect on the WTP. In other words, people who are already experiencing inconveniences from high PM, such as increased expenditure on anti-dust products and deteriorating health, desire UFPs information. People perform various anti-dust activities, such as wearing face masks, refraining from outdoor activities, or using air purifiers, to avoid exposure when PM concentration is high (Cho and Kim, 2019; Noonan, 2014; Saberian et al., 2017; Wells et al., 2012). They use real-time PM information to decide upon such behaviors. Consequently, they believe that accurate information is important to prevent potential risks from PM exposure, and therefore, their WTP for UFPs monitoring and reporting system is high. Additionally, we found that the more reliable the PM information provided by the government, the higher the WTP for a UFPs monitoring and reporting system.

Thus, it is important to accurately measure PM concentration and provide reliable PM information to increase public acceptance of UFPs monitoring and reporting systems. Furthermore, we identified that the higher the level of perception of UFPs, the higher the WTP. Most of our respondents had a low perception of UFPs, but it is expected that the need for UFPs information will increase in future as public perception of UFPs increases. Therefore, policymakers should continuously monitor the public's perception of UFPs, and allocate budgets and establish policies so that the UFPs monitoring and reporting system can be implemented at an appropriate time.

This study is meaningful as the first study to estimate the economic value of UFPs information. However, it also has several limitations. First, the survey was online, which means only the internet savvy could participate. As shown in Table 1, the distribution of sex, age, and income for the respondents was similar to that of the total population, but the respondents' education level was slightly higher than that of the general population. This could be because the internet-literate generally have a relatively higher education level. In general, the WTP among the highly educated is high (Kim et al., 2018; Wang and Zhang, 2009; Wang et al., 2006). We employed an online survey because of COVID-19, but a face-to-face survey should be conducted in future to overcome this limitation and to obtain results that are more accurate. Second, we cannot compare the estimated results of this study with previous studies because to the best of our knowledge, there are no similar studies on the subject and scenarios. Given that concerns about UFPs are increasing, more studies are required to estimate their potential risks and social costs. Third, our respondents had generally low perception of UFPs. We found that respondents with pre-survey knowledge of UFPs placed a higher value on UFPs information. Therefore, there is a possibility that the WTP will change as the perception of UFPs changes in future; therefore, a follow-up research is imperative.

**Funding**

This research did not receive any specific grant from funding agencies in the public, commercial, or not-for-profit sectors.

# References

Akhtar, S., Saleem, W., Nadeem, V.M., Shahid, I., Ikram, A., 2017. Assessment of willingness to pay for improved air quality using contingent valuation method. Glob. J. Environ. Sci. Manag. 3. https://doi.org/10.22034/gjesm.2017.03.03.005

Atkinson, R.W., Kang, S., Anderson, H.R., Mills, I.C., Walton, H.A., 2014. Epidemiological time series studies of PM2.5 and daily mortality and hospital admissions: a systematic review and meta-analysis. Thorax 69, 660–665. https://doi.org/10.1136/thoraxjnl-2013-204492

Bateman, I.J., Langford, I.H., 1997. Non-users' Willingness to Pay for a National Park: An Application and Critique of the Contingent Valuation Method. Reg. Stud. 31, 571–582. https://doi.org/10.1080/00343409750131703

Bishop, R.C., Heberlein, T.A., 1979. Measuring Values of Extramarket Goods: Are Indirect Measures Biased? Am. J. Agric. Econ. 61, 926–930. https://doi.org/10.2307/3180348

Buccolieri, R., Sandberg, M., Di Sabatino, S., 2010. City breathability and its link to pollutant concentration distribution within urban-like geometries. Atmos. Environ. 44, 1894–1903. https://doi.org/10.1016/j.atmosenv.2010.02.022

Burgan, O., Smargiassi, A., Perron, S., Kosatsky, T., 2010. Cardiovascular effects of sub-daily levels of ambient fine particles: a systematic review. Environ. Health 9, 26. https://doi.org/10.1186/1476-069X-9-26

Cesaroni, G., Forastiere, F., Stafoggia, M., Andersen, Z.J., Badaloni, C., Beelen, R., Caracciolo, B., de Faire, U., Erbel, R., Eriksen, K.T., Fratiglioni, L., Galassi, C., Hampel, R., Heier, M., Hennig, F., Hilding, A., Hoffmann, B., Houthuijs, D., Jockel, K.-H., Korek, M., Lanki, T., Leander, K., Magnusson, P.K.E., Migliore, E., Ostenson, C.-G., Overvad, K., Pedersen, N.L., J, J.P., Penell, J., Pershagen, G., Pyko, A.,

Raaschou-Nielsen, O., Ranzi, A., Ricceri, F., Sacerdote, C., Salomaa, V., Swart, W., Turunen, A.W., Vineis, P., Weinmayr, G., Wolf, K., de Hoogh, K., Hoek, G., Brunekreef, B., Peters, A., 2014. Long term exposure to ambient air pollution and incidence of acute coronary events: prospective cohort study and meta-analysis in 11 European cohorts from the ESCAPE Project. BMJ 348, f7412–f7412. https://doi.org/10.1136/bmj.f7412

Chen, R., Hu, B., Liu, Y., Xu, J., Yang, G., Xu, D., Chen, C., 2016. Beyond PM2.5: The role of ultrafine particles on adverse health effects of air pollution. Biochim. Biophys. Acta 1860, 2844–2855. https://doi.org/10.1016/j.bbagen.2016.03.019

Cho, M.E., Kim, M.J., 2019. Residents' Perceptions of and Response Behaviors to Particulate Matter—A Case Study in Seoul, Korea. Appl. Sci. 9, 3660. https://doi.org/10.3390/app9183660

Choi, W., He, M., Barbesant, V., Kozawa, K.H., Mara, S., Winer, A.M., Paulson, S.E., 2012. Prevalence of wide area impacts downwind of freeways under pre-sunrise stable atmospheric conditions. Atmos. Environ. 62, 318–327. https://doi.org/10.1016/j.atmosenv.2012.07.084

Choi, W., Ranasinghe, D., DeShazo, J.R., Kim, J.-J., Paulson, S.E., 2018. Where to locate transit stops: Cross-intersection profiles of ultrafine particles and implications for pedestrian exposure. Environ. Pollut. 233, 235–245. https://doi.org/10.1016/j.envpol.2017.10.055

Cooper, J.C., Hanemann, M., Signorello, G., 2002. One-and-One-Half-Bound Dichotomous Choice Contingent Valuation. Rev. Econ. Stat. 84, 742–750. https://doi.org/10.1162/003465302760556549

Dockery, D.W., Pope, C.A., Xu, X., Spengler, J.D., Ware, J.H., Fay, M.E., Ferris, B.G., Speizer, F.E., 1993. An Association between Air Pollution and Mortality in Six U.S.

Cities. N. Engl. J. Med. 329, 1753–1759. https://doi.org/10.1056/NEJM199312093292401

Fonta, W.M., Ichoku, H.E., Kabubo-Mariara, J., 2010. The Effect of Protest Zeros on Estimates of Willingness to Pay in Healthcare Contingent Valuation Analysis: Appl. Health Econ. Health Policy. 8, 225–237. https://doi.org/10.2165/11530400-000000000-00000

Han, F., Yang, Z., Wang, H., Xu, X., 2011. Estimating willingness to pay for environment conservation: a contingent valuation study of Kanas Nature Reserve, Xinjiang, China. Environ. Monit. Assess. 180, 451–459. https://doi.org/10.1007/s10661-010-1798-4

Hanemann, W.M., 1985. Some Issues in Continuous- and Discrete-Response Contingent Valuation Studies. Northeast. J. Agric. Resour. Econ. 14, 5–13. https://doi.org/10.1017/S0899367X00000702

Hanemann, W.M., 1984. Welfare Evaluations in Contingent Valuation Experiments with Discrete Responses. Am. J. Agric. Econ. 66, 332. https://doi.org/10.2307/1240800

Hart, J.E., Puett, R.C., Rexrode, K.M., Albert, C.M., Laden, F., 2015. Effect Modification of Long-Term Air Pollution Exposures and the Risk of Incident Cardiovascular Disease in US Women. J. Am. Heart Assoc. 4. https://doi.org/10.1161/JAHA.115.002301

HEI, 2013. Understanding the Health Effects of Ambient Ultrafine Particles.

Hu, S., Fruin, S., Kozawa, K., Mara, S., Paulson, S.E., Winer, A.M., 2009. A wide area of air pollutant impact downwind of a freeway during pre-sunrise hours. Atmos. Environ. 43, 2541–2549. https://doi.org/10.1016/j.atmosenv.2009.02.033

IMF, (International Monetary Fund), 2021. Exchange Rate [WWW Document]. URL https://www.imf.org/external/np/fin/data/rms_rep.aspx (accessed 3.22.21).

Istamto, T., Houthuijs, D., Lebret, E., 2014. Multi-country willingness to pay study on road-traffic environmental health effects: are people willing and able to provide a number?

Environ. Health 13, 35. https://doi.org/10.1186/1476-069X-13-35

Jorgensen, B.S., Syme, G.J., 2000. Protest responses and willingness to pay: attitude toward paying for stormwater pollution abatement. Ecol. Econ. 33, 251–265. https://doi.org/10.1016/S0921-8009(99)00145-7

Kim, J.-H., Kim, H.-J., Yoo, S.-H., 2018. Public Value of Enforcing the PM2.5 Concentration Reduction Policy in South Korean Urban Areas. Sustainability 10, 1144. https://doi.org/10.3390/su10041144

Kittelson, D.B., 1998. Engines and nanoparticles. J. Aerosol Sci. 29, 575–588. https://doi.org/10.1016/S0021-8502(97)10037-4

Krinsky, I., Robb, A.L., 1986. On Approximating the Statistical Properties of Elasticities. Rev. Econ. Stat. 68, 715. https://doi.org/10.2307/1924536

Kriström, B., 1997. Spike Models in Contingent Valuation. Am. J. Agric. Econ. 79, 1013–1023. https://doi.org/10.2307/1244440

Kumar, P., Fennell, P., Britter, R., 2008. Measurements of particles in the 5–1000 nm range close to road level in an urban street canyon. Sci. Total Environ. 390, 437–447. https://doi.org/10.1016/j.scitotenv.2007.10.013

Kumar, P., Fennell, P.S., Hayhurst, A.N., Britter, R.E., 2009. Street Versus Rooftop Level Concentrations of Fine Particles in a Cambridge Street Canyon. Bound.-Layer Meteorol. 131, 3–18. https://doi.org/10.1007/s10546-008-9300-3

Kumar, P., Morawska, L., Birmili, W., Paasonen, P., Hu, M., Kulmala, M., Harrison, R.M., Norford, L., Britter, R., 2014. Ultrafine particles in cities. Environ. Int. 66, 1–10. https://doi.org/10.1016/j.envint.2014.01.013

Lee, J., Cho, Y., 2020. Estimation of the usage fee for peer-to-peer electricity trading platform: The case of South Korea. Energy Policy 136, 111050. https://doi.org/10.1016/j.enpol.2019.111050

Lee, Y.J., Lim, Y.W., Yang, J.Y., Kim, C.S., Shin, Y.C., Shin, D.C., 2011. Evaluating the PM damage cost due to urban air pollution and vehicle emissions in Seoul, Korea. J. Environ. Manage. 92, 603–609. https://doi.org/10.1016/j.jenvman.2010.09.028

Lewis, A., Carslaw, D., Moller, S.J., 2018. Ultrafine Particles (UFP) in the UK. United Kingdom: Defra (Department for Environment, Food & Rural Affairs).

Li, T., Zhang, Y., Wang, J., Xu, D., Yin, Z., Chen, H., Lv, Y., Luo, J., Zeng, Y., Liu, Y., Kinney, P.L., Shi, X., 2018. All-cause mortality risk associated with long-term exposure to ambient PM2·5 in China: a cohort study. Lancet Public Health 3, e470–e477. https://doi.org/10.1016/S2468-2667(18)30144-0

Liang, C.-S., Duan, F.-K., He, K.-B., Ma, Y.-L., 2016. Review on recent progress in observations, source identifications and countermeasures of PM2.5. Environ. Int. 86, 150–170. https://doi.org/10.1016/j.envint.2015.10.016

Ligus, M., 2018. Measuring the Willingness to Pay for ImprovedAir Quality: A Contingent Valuation Survey. Pol. J. Environ. Stud. 27, 763–771. https://doi.org/10.15244/pjoes/76406

Liu, J.-Y., Hsiao, T.-C., Lee, K.-Y., Chuang, H.-C., Cheng, T.-J., Chuang, K.-J., 2018. Association of ultrafine particles with cardiopulmonary health among adult subjects in the urban areas of northern Taiwan. Sci. Total Environ. 627, 211–215. https://doi.org/10.1016/j.scitotenv.2018.01.218

Marini, S., Buonanno, G., Stabile, L., Avino, P., 2015. A benchmark for numerical scheme validation of airborne particle exposure in street canyons. Environ. Sci. Pollut. Res. 22, 2051–2063. https://doi.org/10.1007/s11356-014-3491-6

ME, (Ministry of Environment), 2021. 2021 Ministry of Environment Budget. Ministry of Environment.

ME, (Ministry of Environment), 2020a. Monthly Report of Air Quality in Korea, October

2020. Ministry of Environment.

ME, (Ministry of Environment), 2020b. Annual Report of Air Quality in Korea, 2019. Ministry of Environment.

Meyerhoff, J., Liebe, U., 2010. Determinants of protest responses in environmental valuation: A meta-study. Ecol. Econ. 70, 366–374. https://doi.org/10.1016/j.ecolecon.2010.09.008

Min, J.W., 2019. Survey on public perception of particulate matter. Mon. Econ. Rev. 833.

MOIS, (Ministry of the Interior and Safety), 2021. URL https://jumin.mois.go.kr/index.jsp# (accessed 3.11.21).

Morawska, L., Ristovski, Z., Jayaratne, E.R., Keogh, D.U., Ling, X., 2008. Ambient nano and ultrafine particles from motor vehicle emissions: Characteristics, ambient processing and implications on human exposure. Atmos. Environ. 42, 8113–8138. https://doi.org/10.1016/j.atmosenv.2008.07.050

Mwebaze, P., Marris, G.C., Brown, M., MacLeod, A., Jones, G., Budge, G.E., 2018. Measuring public perception and preferences for ecosystem services: A case study of bee pollination in the UK. Land Use Policy 71, 355–362. https://doi.org/10.1016/j.landusepol.2017.11.045

Noonan, D.S., 2014. Smoggy with a Chance of Altruism: The Effects of Ozone Alerts on Outdoor Recreation and Driving in Atlanta: Effects of Ozone Alerts. Policy Stud. J. 42, 122–145. https://doi.org/10.1111/psj.12045

Oerlemans, L.A.G., Chan, K.-Y., Volschenk, J., 2016. Willingness to pay for green electricity: A review of the contingent valuation literature and its sources of error. Renew. Sustain. Energy Rev. 66, 875–885. https://doi.org/10.1016/j.rser.2016.08.054

Ohlwein, S., Kappeler, R., Kutlar Joss, M., Künzli, N., Hoffmann, B., 2019. Health effects of ultrafine particles: a systematic literature review update of epidemiological evidence.

Int. J. Public Health 64, 547–559. https://doi.org/10.1007/s00038-019-01202-7

Penttinen, P., Timonen, K.L., Tiittanen, P., Mirme, A., Ruuskanen, J., Pekkanen, J., 2001. Ultrafine particles in urban air and respiratory health among adult asthmatics. Eur. Respir. J. 17, 428–435. https://doi.org/10.1183/09031936.01.17304280

Saberian, S., Heyes, A., Rivers, N., 2017. Alerts work! Air quality warnings and cycling. Resour. Energy Econ. 49, 165–185. https://doi.org/10.1016/j.reseneeco.2017.05.004

Schraufnagel, D.E., 2020. The health effects of ultrafine particles. Exp. Mol. Med. 52, 311–317. https://doi.org/10.1038/s12276-020-0403-3

Song, S., Lee, K., Lee, Y.-M., Lee, J.-H., Lee, S.I., Yu, S.-D., Paek, D., 2011. Acute health effects of urban fine and ultrafine particles on children with atopic dermatitis. Environ. Res. 111, 394–399. https://doi.org/10.1016/j.envres.2010.10.010

Stafoggia, M., Schneider, A., Cyrys, J., Samoli, E., Andersen, Z.J., Bedada, G.B., Bellander, T., Cattani, G., Eleftheriadis, K., Faustini, A., Hoffmann, B., Jacquemin, B., Katsouyanni, K., Massling, A., Pekkanen, J., Perez, N., Peters, A., Quass, U., Yli-Tuomi, T., Forastiere, F., UF&HEALTH Study Group, 2017. Association Between Short-term Exposure to Ultrafine Particles and Mortality in Eight European Urban Areas. Epidemiol. Camb. Mass 28, 172–180. https://doi.org/10.1097/EDE.0000000000000599

Strazzera, E., Scarpa, R., Calia, P., Garrod, G.D., Willis, K.G., 2003. Modelling zero values and protest responses in contingent valuation surveys. Appl. Econ. 35, 133–138. https://doi.org/10.1080/00036840022000015900

Sun, C., Yuan, X., Xu, M., 2016. The public perceptions and willingness to pay: from the perspective of the smog crisis in China. J. Clean. Prod. 112, 1635–1644. https://doi.org/10.1016/j.jclepro.2015.04.121

Tolunay, A., Başsüllü, Ç., 2015. Willingness to Pay for Carbon Sequestration and Co-Benefits

of Forests in Turkey. Sustainability 7, 3311–3337. https://doi.org/10.3390/su7033311

Ünver, H., 2014. Explaining Education Level and Internet Penetration by Economic Reasoning - Worldwide Analysis from 2000 through 2010. Int. J. Infonomics 7, 898–912.

Vlachokostas, C., Achillas, C., Slini, T., Moussiopoulos, N., Banias, G., Dimitrakis, I., 2011. Willingness to pay for reducing the risk of premature mortality attributed to air pollution: a contingent valuation study for Greece. Atmospheric Pollut. Res. 2, 275–282. https://doi.org/10.5094/APR.2011.034

Wang, X.J., Zhang, W., Li, Y., Yang, K.Z., Bai, M., 2006. Air Quality Improvement Estimation and Assessment Using Contingent Valuation Method, A Case Study in Beijing. Environ. Monit. Assess. 120, 153–168. https://doi.org/10.1007/s10661-005-9054-z

Wang, Y., Zhang, Y.-S., 2009. Air quality assessment by contingent valuation in Ji'nan, China. J. Environ. Manage. 90, 1022–1029. https://doi.org/10.1016/j.jenvman.2008.03.011

Wells, E.M., Dearborn, D.G., Jackson, L.W., 2012. Activity Change in Response to Bad Air Quality, National Health and Nutrition Examination Survey, 2007–2010. PLoS ONE 7, e50526. https://doi.org/10.1371/journal.pone.0050526

Wu, H., Huang, J., Zhang, C.J.P., He, Z., Ming, W.-K., 2020. Facemask shortage and the novel coronavirus disease (COVID-19) outbreak: Reflections on public health measures. EClinicalMedicine 21, 100329. https://doi.org/10.1016/j.eclinm.2020.100329

Yuan, S., Wang, J., Jiang, Q., He, Z., Huang, Y., Li, Z., Cai, L., Cao, S., 2019. Long-term exposure to PM2.5 and stroke: A systematic review and meta-analysis of cohort studies. Environ. Res. 177, 108587. https://doi.org/10.1016/j.envres.2019.108587

**Table 1. Demographic characteristics of survey respondents**

| Characteristic | Group | Survey respondents | | General population [a] |
|---|---|---|---|---|
| | | Number of respondents (n = 1040) | Ratio (%) | Ratio (%) |
| Sex | Male | 553 | 53.17 | 50.88 |
| | Female | 487 | 46.83 | 49.12 |
| Age | 20–29 | 184 | 17.69 | 17.76 |
| | 30–39 | 217 | 20.87 | 17.96 |
| | 40–49 | 245 | 23.56 | 21.85 |
| | 50–59 | 244 | 23.46 | 23.15 |
| | 60–69 | 150 | 14.42 | 19.29 |
| Region | Metropolitan City | 477 | 45.87 | 44.27 |
| | Non-metropolitan City | 563 | 54.13 | 55.73 |
| Education level (graduation) | Lower than high school | 142 | 13.65 | 50.00 |
| | Higher than University | 898 | 86.35 | 50.00 |
| Average monthly income per household (KRW 10,000) | Under 299 | 278 | 26.73 | 27.69 |
| | 300–399 | 157 | 15.10 | 14.65 |
| | 400–499 | 160 | 15.38 | 14.11 |
| | 500–699 | 236 | 22.69 | 22.57 |
| | Over 700 | 209 | 20.10 | 20.98 |
| Average monthly expense for anti-dust products (KRW 1,000) | Under 5 | 129 | 12.40 | 29.20 |
| | 6–10 | 181 | 17.40 | 25.70 |
| | 11–20 | 310 | 29.81 | 18.70 |
| | 21–30 | 232 | 22.31 | |
| | 31–40 | 117 | 11.25 | 15.70 |
| | 41–50 | 33 | 3.17 | |
| | Over 50 | 38 | 3.65 | 10.70 |

[a] Sex, age, and region are for the 20–69 years old population in 2020 (Source: MOIS, 2021); Education level is based on 2019 (Source: OECD, 2020); Average monthly income per household is for the entire population in 2019 (Source: KOSIS, 2021);Average monthly expense for anti-dust products is a survey sample of 1,008 from 2019 (Source: Min, 2019)

**Table 2. Respondents' cognition regarding particulate matter**

| Cognition | Question | Response | Number of respondents | Ratio (%) |
|---|---|---|---|---|
| Seriousness | I think the PM problem in Korea is serious. | Very unlikely | 0 | 0.00 |
| | | Unlikely | 2 | 0.19 |
| | | Average | 106 | 10.19 |
| | | Likely | 591 | 56.83 |
| | | Very likely | 341 | 32.79 |
| Adverse health effect | I think PM has a negative effect on health. | Very unlikely | 21 | 2.02 |
| | | Unlikely | 189 | 18.17 |
| | | Average | 483 | 46.44 |
| | | Likely | 296 | 28.46 |
| | | Very likely | 51 | 4.90 |
| Reliability of real-time PM data | I think real-time PM concentration information provided by government institutes is reliable. | Very unlikely | 33 | 3.17 |
| | | Unlikely | 106 | 10.19 |
| | | Average | 383 | 36.83 |
| | | Likely | 426 | 40.96 |
| | | Very likely | 92 | 8.85 |
| Perception of UPFs | I think I am aware of UFPs before participating in this survey. | Very unlikely | 120 | 11.54 |
| | | Unlikely | 367 | 35.29 |
| | | Average | 349 | 33.56 |
| | | Likely | 162 | 15.58 |
| | | Very likely | 42 | 4.04 |

**Table 3. Distribution of respondents in the survey**

| Bid amount (KRW) | | Upper bid is given as an initial bid | | | | | Lower bid is given as an initial bid | | | | |
|---|---|---|---|---|---|---|---|---|---|---|---|
| Lower | Upper | "yes" | "no-yes" | "no-no-yes" | "no-no-no" | Total | "yes-yes" | "yes-no" | "no-yes" | "no-no" | Total |
| 1000 | 2000 | 27 | 8 | 3 | 15 | 53 | 19 | 11 | 8 | 9 | 47 |
| 2000 | 3000 | 24 | 6 | 20 | 14 | 64 | 12 | 3 | 10 | 12 | 37 |
| 3000 | 4000 | 18 | 5 | 13 | 9 | 45 | 24 | 7 | 16 | 12 | 59 |
| 4000 | 5000 | 28 | 3 | 14 | 18 | 63 | 13 | 8 | 17 | 10 | 48 |
| 5000 | 7000 | 28 | 6 | 13 | 12 | 59 | 8 | 9 | 17 | 9 | 43 |
| 7000 | 9000 | 23 | 4 | 17 | 8 | 52 | 16 | 3 | 16 | 21 | 56 |
| 9000 | 11000 | 21 | 7 | 19 | 9 | 56 | 12 | 8 | 14 | 14 | 48 |
| 11000 | 14000 | 15 | 11 | 16 | 16 | 58 | 13 | 11 | 11 | 14 | 49 |
| 14000 | 17000 | 10 | 7 | 22 | 8 | 47 | 10 | 5 | 25 | 12 | 52 |
| 17000 | 20000 | 9 | 4 | 26 | 9 | 48 | 8 | 6 | 24 | 18 | 56 |
| Total (%) | | 203 (37.2) | 61 (11.2) | 163 (29.9) | 118 (21.7) | 545 (100) | 135 (27.3) | 71 (14.3) | 158 (31.9) | 131 (26.5) | 495 (100) |

**Table 4. Reasons for zero WTP**

| Reasons for zero WTP | Number of respondents | Ratio (%) |
| --- | --- | --- |
| Cannot afford to pay the additional income tax | 43 | 17.27 |
| Should be covered by the existing income tax | 90 | 36.14 |
| Not enough information to judge | 38 | 15.26 |
| Not important enough to prioritize | 43 | 17.27 |
| Not interested | 28 | 11.24 |
| Others | 7 | 2.81 |
| Total of reasons mentioned | 249 | 100 |

**Table 5. Estimation results**

| Variables | Model 1 | Model 2 | Model 3 |
|---|---|---|---|
| Constant | 0.991 (13.72) ** | -0.137 (-0.45) | -3.711 (-6.64) ** |
| Bid [c] | 0.181 (25.38) ** | 0.183 (25.38) ** | 0.197 (25.37) ** |
| Sex | | 0.024 (0.21) | 0.042 (0.35) |
| Age | | 0.012 (2.67) ** | 0.009 (1.98) * |
| Income | | 0.068 (1.70) | 0.020 (0.47) |
| Anti-dust expense | | 0.119 (2.99) ** | 0.080 (1.87) |
| Seriousness | | | 0.208 (2.17) * |
| Adverse health effect | | | 0.171 (2.28) * |
| Reliability of real-time PM data | | | 0.568 (8.04) ** |
| Perception of UFPs | | | 0.245 (3.97) ** |
| Spike | 0.271 (18.96) ** | 0.267 (18.79) ** | 0.253 (17.61) ** |
| Log-likelihood | -1459.17 | -1448.79 | -1393.84 |
| Wald statistic (p-value) | 756.08 (0.000) | 782.10 (0.000) | 815.35 (0.000) |
| Mean WTP (KRW) | 7222.55 (28.04) * | 7196.33 (27.97) ** | 6958.55 (28.55) ** |
|   95% Confidence interval | 6716.89–7744.53 | 6704.69–7712.78 | 6498.26–7443.92 |
|   99% Confidence interval | 6589.88–7917.09 | 6565.50–7870.22 | 6354.49–7614.11 |

Notes: Values in bracket are t-statistics except Wald statistic.

* Statistically significant at the 5% level. ** Statistically significant at the 1% level.

**Table 6. Summary of the scenario and estimated WTP of the previous studies using the contingent valuation method**

| Study | Survey year | Country | Scenario | Payment vehicle | Reported WTP | Value transfer to 2020 USD [a] |
|---|---|---|---|---|---|---|
| Akhtar et al. (2017) | 2016 | Lahore, Pakistan | Reduction of atmospheric contamination by 50% | Voluntary payment | USD 118 [b] | 156.65 |
| Istamto et al. (2014) | 2010 | Finland, Germany, Netherlands, Spain, United Kingdom | Increased life expectancy by six months because of 50% reduction in air pollution emissions by 2030 | Voluntary payment | EURO 82 [c] | 73.77 |
| Kim et al. (2018) | 2017 | Korea | Reinforcement of PM2.5 concentration reduction policy | Additional income tax | USD 4.97 [b] | 5.09 |
| Lee et al. (2011) | 2006 | Seoul, Korea | Reduced risk of mortality due to air pollution from motor vehicles by 5/1,000 (or 1/1,000) over 10 years | Voluntary payment | USD 242.4 [c] | 318.63 |
| Ligus (2018) | 2015 | Poland | Avoiding six risks: mortality, morbidity, damage to historical buildings and monuments, material damage, visibility loss, and ecosystem damage, through air quality improvement | Additional electricity bills | PLN 235.25 [c] | 69.52 |
| Sun et al. (2016) | 2013 | China | Reduced premature death and respiratory disease through the air quality improvement project | Voluntary payment | RMB 382.6 [b] | 67.82 |
| Vlachokostas et al. (2011) | 2009 | Thessaloniki, Greece | Increased life expectancy by one year because of air quality improvement | Additional green tax | EURO 920 [c] | 797.19 |
| Wang and Zhang (2009) | 2006 | Jinan, China | Applying a strict national air quality standard | Voluntary payment | CNY 100 [c] | 22.29 |

| Wang et al. (2006) | 1999 | Beijing, China | 50% reduction of harmful substances | Voluntary payment | CNY 143 [b] | 34.75 |

[a] The WTP in 2020 USD is calculated by the consumer price index (CPI) of the survey year and 2020 and the average exchange rate on 2020/12/23. The CPI is from IMF (2020), and the average exchange rate is from IMF (2021).

[b] per household per year; [c] per person per year